\documentstyle[12pt]{article}
\def\kms{km s$^{-1}$}
\def\Msun{M_{\odot \hskip-5.2pt \bullet}}
\def\r{\hangindent=1pc  \noindent}
\def\kms{km s$^{-1}$}

\def\ta{$T_{\rm A}^*$}
\def\Msun{M_{\odot \hskip-5.2pt \bullet}}
\def\co{$^{12}$CO($J=1-0$)}

\def\tfr{{Tully-Fisher relation}}

\begin{document}

\title{CO vs HI ROTATIONS OF GALAXIES}

\author{{Yoshiaki \sc Sofue}\\
{\it Institute of Astronomy, University of Tokyo, Mitaka, Tokyo 181}\\
{\it Email: sofue@mtk.ioa.s.u-tokyo.ac.jp}}

\maketitle

\abstract{Since molecular gas is distributed in the inner disk
of galaxies while HI in the outer disk,  
CO emission can trace the inner rotation curves more accurately than HI.
We have derived most-completely sampled rotation curves by combining
HI and CO position-velocity diagrams.
We show that inner rotation curves have  a steep 
increase within the central few hundred-pc region.
The inner behavior of rotation can be fitted by the Miyamoto-Nagai potential, 
only if a central mass component with scale radius of 100 -- 200 pc and a 
mass of $\sim 3-5\times 10^{9}\Msun$ is assumed in addition to the usual bulge.
This implies that the central $\sim 100$-pc region has a stellar mass density
 as high as several hundred $\Msun{\rm  pc}^{-3}$, orders-of-magnitudes 
higher than that for a single bulge and disk (several $\Msun {\rm pc}^{-3}$).
Such a concentric multi-component structure within a bulge 
(``bulge in bulge'') may have crucial implication for the dynamical 
evolution of the central regions of galaxies.
Since the molecular gas distribution is closely correlated with 
optical disk, CO line can be used for the line width-luminosity (Tully Fisher)
relation in addition to (instead of) the HI line.
We show that the CO-line width-luminosity relation can be a powerful
routine method to derive distances of cosmologically-remote galaxies
and the Hubble parameter for  a significantly large volume of the universe.
By applying the CO-line width-luminosity relation to CO-line data
by large-aperture mm-wave telescope,  we have derived distances
to galaxies at $cz \sim 20,000$.
In order to measure the inclination and magnitudes,
we performed optical imaging observations of thus CO-detected galaxies
using the CFHT 3.6-m telescope at high resolution.
The radio and optical data have been combined to derive the distances
to the galaxies. We further derived the Hubble ratio for these galaxies
by applying the CO-line width-luminosity relation.
\footnote{ Contribution at  a 
 Japanese-German JSPS-DFG-Seminar
at Inst Theor Astrophys
Uni Heidelberg 06 - 10 November 1995:
"Formation and Evolution of Galaxies"
(eds. W.Duschl \& N.Arimoto)}

{\bf Key words} {Galaxies: kinematics -- Galaxies: rotation -- 
Galaxies: distances -- Galaxies: ISM}

}

\section{Inner CO vs Outer HI Rotation Curves}

\subsection{Most-Completely Sampled Rotation Curves} 
\label{title}

Rotation curves (RC) are the principal tool to derive the distribution
of mass in galaxies, and have been investigated extensively based
on optical (H$\alpha$) and HI-line observations along
the major axes of galaxies (Rubin et al 1980, 1982; Persic et al 1995) and
the Galactic plane (Burton and Gordon 1978; Clemens 1985).
Recently, we obtained most-completely-sampled rotation curves for many nearby 
galaxies by combining high-resolution CO and HI position-velocity diagrams 
with a particular concern with central behavior of RC (Sofue 1996a).
It was shown that RC for galaxies in this study rises very steeply 
within 100 -- 200 pc region of the center (Fig. 1).
Among them, the inner RC of NGC 891, NGC 3079, and NGC 6946 were shown
to have a sharp central peak, very similar to that observed for the Milky Way.
Such a steep rising has been not well detected in the current optical
and HI RC analyses.

We show that the innermost RC can be well reproduced by a mass model 
in which a fourth mass component is present within the central
bulge in addition to the usual bulge, disk and massive halo (Sofue 1996b).
We discuss the implication of the four-mass component model and 
a ``bulge-in-bulge'' structure for the dynamical evolution of the central
 region of galaxies.

\subsection{Fitting by Four-Mass Component Potential: Bulge in Bulge}

We try to fit the RC by the Miyamoto-Nagai (MN) (1975) potential, which
is  most suitable to represent an axisymmetric mass distribution 
comprising  multiple components satisfying  the Poisson's equation.
The  MN potential with $n$ mass components is expressed as 
$$ 
\Phi= \sum_{i=1}^n  {GM_i  \lbrack R^2 + \lbrace a_i 
+ (z^2+b_i^2)^{1/2} \rbrace ^2} \rbrack  ^{-1/2}, 
$$
Here, $R$ denotes  the distance from the rotation axis, $z$ is the height 
from the galactic plane, and $M_i$, $a_i$ and $b_i$ are the mass, scale 
radius, and scale thickness, respectively, of the $i$-th mass component.
For a spherical mass distribution such as for a bulge, 
we have $a_i=0$, and $b_i$  becomes equal to the scale radius of the sphere.
The rotation velocity is calculated by
$$
V_{\rm rot} = \left( R {\partial \Phi / \partial R} \right)^{1/2}. 
$$
Miyamoto and Nagai (1975) have assumed two components ($n=2$) in order
to describe the bulge + disk structure of the Galaxy.
In order to fit the flat rotation at $R\sim 10 - 20$ kpc, an extended 
massive halo has to be introduced, and  a three-component model ($n=3$) 
is widely  used, assuming a bulge, disk, and massive halo.

After a trial of fitting to the RC of the central few hundred pc region, 
it turned out that the usual three-component model is not sufficient 
to fit the steep central peak.
It is necessary to introduce a fourth component which represents
a more compact component in addition to the usual bulge.

In  the first panel of Fig. 1 we show a RC of the Galaxy calculated by this 
four-component model (thin full line), and the observed RC from  Clemens 
(1985) is superposed by a thick line.
Dashed lines indicate RC corresponding to individual components.
 The observed RC is fitted well by a model with:
(1) A massive halo of $M_1=5\times 10^{11}\Msun$ and scale radius of 
$a_1=b_1=15$ kpc in order to fit the flat part of the RC in the
outer Galaxy;
(2) A  disk of $M_2=1.5\times 10^{11}\Msun$ with scale radius of
 $a_2=6$ kpc and  thickness $b_2=0.5$ kpc. Hereafter, we fix $b_2=0.5$ kpc
for all galaxies, since RC is not sensitive to $b_2$ if $a_2>>b_2$.
(3) A spherical ($a_3=0$) bulge of $M_3=1.5\times10^{10}\Msun$ 
and $b_3=0.9$ kpc radius; and
(4) A spherical ($a_4=0$) central component of $M_4=5.5\times 10^{9}\Msun$ 
of a $b_4=150$ pc scale radius. 
We stress that this  central component is more extended
than the circum-nuclear mass condensation within $R \sim 50 $-pc,
where the stellar density increases toward the nucleus roughly obeying 
a power law ( Genzel and Townes 1987).
In order to reproduce this innermost mass distribution, a fifth or 
more additional components with smaller scale radii would be required.

The RC for the other nine galaxies are also fitted by this model.
Fig. 1 shows thus-calculated RC  compared with the observations.
Generally, all the RC are well reproduced by the model.
Particularly, the sharp central peaks observed for the Galaxy, NGC 891, NGC 
3079 and NGC 6946 are well fitted by the 4-th component with similar 
parameters. Flat valleys of RC at $R\sim$  1 to 2 kpc observed 
for NGC 6946 and NGC 3079 are also well reproduced by the present model,
which are  difficult to fit by any three-component model.
Galaxies like  NGC 253 and IC 342 have no sharp peak of RC, but show a steep
rising up near the center followed by a sudden turn-off to a  flat RC.
This behavior can be well reproduced by the existence of the 4-th component.
Only one case, the outer decline of RC of M51, was difficult 
to fit with the present model:  The RC declines  suddenly  
beyond $R=\sim 9$ kpc, whereas it is almost  flat inside. 
Such a sudden decline of RC cannot be fitted by any simple gravitational 
potential even without the halo component,  and might be due to some 
non-circular motion induced by the tidal interaction with the close companion.
Table 1 summarizes the best-fit parameters for the observed galaxies.
Errors of fitted parameters for the inner three components ($i=2,3,4$)
are within $\pm 15$\% as eye-estimated during trials of fitting.
Errors for the halo components are $\sim\pm 25$\% for galaxies
with RC observed at 15 to 20 kpc, while errors are larger, $\sim50$\%,
for galaxies with RC available only at $R<10-15 $kpc like NGC 253.

The density corresponding to  the potential is given by
$\rho=(1/4\pi G) \Delta \Phi$.
The central density for each component $\rho_i^0$ can be calculated for
$(R,z)=(0,0)$, and we list the values in Table 1.
The 4-th component has a mass density at the center as high as 
$\rho_4^0\sim 10^2 - 10^3\Msun$ pc$^{-3}$, orders of magnitudes higher than 
that estimated from a single-bulge + disk model  
($\rho_3^0\sim 10 \Msun$ pc$^{-3}$).
The density of a spherical component varies with radius $r$ 
as  $\rho=\rho_i^0[(r/b_i)^2+1]^{-2.5}$ with $\rho_i^0=(3M/4\pi) b_i^{-3}$.
The surface density of the spherical component as projected on the sky varies
 as $\sigma_i(r) \simeq \sigma_i^0[(r/b_i)^2+1]^{-2}$ with 
$\sigma_i^0=M/\pi b_i^2$.
We emphasize that the high central density due to the 4th component
superposed on the 3rd component
cannot be obtained by an $e^{-(r/b_i)^{1/4}}$ law for a single bulge.

\begin{figure}
\vspace{20cm}
\caption{Model RC of galaxies as calculated for a Miyamoto-Nagai potential 
with four mass components (thin line) fitted to the observations (thick line).
}
\end{figure}

\begin{table*}[t]
\small
\caption{
 Fitted parameters by the four-mass-component MN potential.
$M_i$ in $10^{11}\Msun$; $a_i$, $b_i$ in kpc; $\rho_i^0$
in $\Msun$ pc$^{-3}$; $ \sigma_i^0$ in $10^3 \Msun$ pc$^{-2}$.
}

\begin{tabular}{lcccccccccccccc}
\hline\hline \\ [-5pt]
 &MW&N253 &IC342 & N891 &N1808 & N3079& N4565&M51&N5907&N6946 \\
 & Sb & Sc & Sc & Sb & Sbc & Sc & Sb &Sc &Sc &Sc  \\
\\
\hline \\ [-6pt]

$M_1$ (Halo)& 5.0& 2.0 & 3.0 &2.0 & 0    &3     & 5& 0 &3.5 &3.0 \\
$a_1=b_1$ & 15     & 15 & 15& 20  & ---    &18  & 15 & ---    &15  &20 \\
$\rho_1^0$ & .0035& .0003&.0020& .0006 & 0& .0012& .0035& 0& .0025&.0009 \\
\\
$M_2$ (Disk)&1.5 & 1.1 & 1.0&1.4& 0.80  &1.50   & 2.2 & 1.3    &2.5 &1.65 \\
$a_2(b_2=0.5)$  & 6.0 & 5.0& 5.5&   5.0 &  4.2   &5.0   & 7.0 & 4.7&7.5 &6..0\\
$\rho_2^0$ & 2.0& 2.1& 1.9& 2.7& 2.2& 2.9& 2.1& 2.8 & 2.1& 2.2\\
\\
$M_3$ (Bulge)&0.15& 0.16& 0.14&0.12&0.10  &0.10 & 0.25 & 0.30  &0.25 & 0.10\\
$b_3(a_3=0)$& 0.9& 1.0& 1.2&  0.8   & 1.0  &1.10& 1.0  & 1.4 &1.5   & 1.10\\
$\rho_3^0$ & 5.0& 3.8& 1.9& 5.5& 2.4& 1.8& 6.0 & 2.6 & 1.8& 1.8\\
$\sigma_3^0 $ & 5.9 & 5.1 & 3.1& 6.0 & 3.2 & 2.6 & 8.0 & 4.9 & 3.5 & 2.6 \\
\\
$M_4$ (Cen.)& 0.055& 0.035& 0.015& 0.05& 0.055 &0.08 & 0.05&0.08&0.05 &0.040\\
$b_4(a_4=0)$& 0.15& 0.16 & 0.15& 0.16 & 0.22 &0.16 & 0.20 &0.25 & 0.25& 0.12\\
$\rho_4^0$ & 390 & 200& 110& 290& 120& 470& 290& 120& 80& 550\\
$\sigma_4^0$ & 78 & 44 & 21 & 62 & 36 & 99 & 40 & 41 & 26 & 88 \\
\\
\hline \\ [-6pt]

\end{tabular}
\end{table*}

\subsection{Implication for Dynamics in the Central Regions}

\subsubsection{Effect of a Bar}

The ``rotation curves'' are obtained as loci of the highest-velocity 
envelopes in position-velocity plots along the major axes of galaxies.
Therefore, non-circular motion due to a bar or oval potential would 
be superposed on the circular motion.
If bar-shocked gas dominates in the CO emission,
and if the bar axis is perpendicular to the line of sight, the radial 
velocity will be almost equal to the pattern speed of bar in
rigid rotation, much smaller than the circular velocity
(e.g., S$\phi$rensen et al 1976).
On the other hand, if the bar is parallel to the line of sight, 
the velocity will be even larger than circular speed.
The fact that all galaxies studied here show a steep rise-up near the center 
should be taken as an indirect evidence for a weak (or no) bar-shock,
because the probability of looking at a bar parallel is small.
We also mention that the CO gas in the central 150 pc of our Galaxy is nearly
in circular rotation, and non-circular motion is at most $\sim 40$ \kms 
(Sofue 1995).
Velocity fields observed for the central regions of NGC 6946 and IC 342, 
known for a bar of molecular-gas in the center (e.g., Ishizuki et al. 1990),
also indicate that non-circular velocities are $\sim \pm 40$ \kms.

Weak-bar approximation within the central few hundred parsecs would 
be reasonable also from another point of view:
According to the present analysis, the potential in the central 
a few hundred-pc region is dominated by the spherical condensation of mass,
 whose density and total mass are  orders of magnitudes greater  than 
the density and mass due to the disk component.
Therefore, even if a bar instability  is induced in the disk, it may be easily
stabilized by the massive spherical mass.

Hence, the circular-motion approximation assumed here would be reasonable
in so far as the zero-th-order mass distribution in the
central region of galaxies is concerned.

\subsubsection{Higher Central Density in Earlier Hubble Type Galaxies}

It is interesting to examine if the parameters listed in Table 1, particularly
those related to bulges, are correlated with the Hubble type.
In Fig. 2 we show some correlations.
The central density of the disk component ($\rho_2^0$) is almost constant.
It is conspicuous that bulges for Sb galaxies have much 
greater central density than Sc.
This indicates that the morphologically ``smaller-bulge'' for later-type
galaxies is due to lower density of bulge, but not due to difference 
in scale radii.
In fact, bulge radius $b_3$ and the ratio $b_3/a_2$ rather increase toward Sc.
The central density $\rho_4^0$  also show a similar behavior
to $\rho_3$, except for NGC 3079 and NGC 6946.
The high density for NGC 3079 might be related to formation
of the central compact object associated with nuclear activity and jet.
Among the other parameters, $M_2$ is well correlated with $a_2$,
and $M_3$ and  $M_4$  are also correlated with $a_3$ and 
$a_4$, respectively.

\begin{figure}
\vspace{10cm}
\caption{The central densities plotted against galaxy types.
Sb galaxies have higher density, corresponding to their 
larger  (more massive)bulges, than Sc galaxies.
}
\end{figure}

\subsubsection{Evolution of Bulges and Central Regions}

The concentric ``bulge-in-bulge'' structure suggests that the central 
mass condensation during the dynamical evolution of galaxies was not 
single-fold, but it may have occurred recurrently under different conditions.
In addition to a bulge formation during the initial contraction (e.g., 
Eggen et al. 1962), recurrent accretion of gas and subsequent starbursts
due to barred-shocks (Noguchi 1988; Wada and Habe 1992) would have resulted 
in formation of a denser stellar component within the bulge.
According to a viscous-disk contraction model with on-going starformation
(Saio and Yoshii 1990), a central density excesses over an exponential disk
and a peaked RC at a several-hundred-pc radius is obtained.
It is likely that the viscosity time scale in the central 100 -- 200 pc
is much shorter than that in the disk due to stronger shear motion, 
while the starformation time scale is similar.
This may result in a higher-density stellar condensation, and would account
for the multi-bulge structure.

\vskip 10mm

\section{CO-Line Width - Luminosity Relation}

\subsection{CO-Line Tully Fisher Relation}

The HI-line width - luminosity relation (\tfr)
is one of the most powerful tools to measure 
distances to galaxies (Tully \& Fisher 1977; Aaronson et al. 1986;  Pierce 
\& Tully 1988; Kraan-Korteweg et al. 1988;  Fouqu{\'e} et al. 1990; 
Fukugita et al. 1991).
However, distances to galaxies so far reached by HI observations are limited
to around 100 Mpc, or $cz \sim 10,000$ to 15,000 \kms\ even
with the use of the world-largest telescopes (Sch\"oniger and Sofue 1993).
We have no routine method to determine distances to galaxies beyond
this distance, at which  angular resolution of a few arc minutes
in HI observations becomes too large to resolve individual 
galaxies in a cluster. 
Interferometers like VLA are not useful for the purpose 
because of the limited number of spectral channels (velocity resolution).
Furthermore,  red-shifted HI frequency results in
increases in beam size as well as in interferences, 
which also makes resolution of distant cluster galaxies difficult.
Moreover, HI line profiles are easily disturbed by interactions among galaxies,
which is inevitable in the central region of a cluster, causing 
uncertainty in the HI line profiles for the \tfr.
On the other hand, molecular gas is tightly correlated with the luminous 
stellar disk, so that it is less affected by the tidal interaction.
The molecular gas is distributed enough to a radius
of several to ten kpc, so that the integrated line profiles manifest the 
maximum velocity part of the rotation curve (Sofue 1992).

Molecular-line observations at millimeter wavelengths, particularly in the
CO-line emission at 115 GHz, can be achieved with  much sharper beams.
Therefore,  we will be able to resolve individual member galaxies
in a cluster more easily, which makes it possible to avoid 
contamination by other member galaxies in a beam.
Moreover, the larger is the redshift of an object, the lower becomes the
CO frequency, which results in a decrease in the system noise temperature
due to the atmospheric O$_2$ emission near 115 GHz:
the more distant is a galaxy, the lower becomes the noise temperature.

Dickey and Kazes (1992) have addressed the use of CO line widths
instead of and/or in supplement to HI observations, and proposed the
CO-line Tully-Fisher relation as an alternative to the HI \tfr.
The CO-line width - luminosity relation has been established for the local 
distance calibrators and tens of nearby galaxies (Sofue 1992;  Sch{\"o}niger 
and Sofue 1993) (Fig. 3).
In these works, we have have shown that the CO-line measurements can be 
used as  an alternative  to HI by deriving a good linear correlation between 
CO and HI linewidths for the galaxies in the sample.
Only a disadvantage of the use of CO line would be the sensitivity, 
particularly for distant galaxies.
Actually, we need a few mK rms data for line-width measurements
at a velocity resolution of 10 \kms\ for normal galaxies beyond 
$cz \sim 10,000$ \kms, which requires long integration times.
Such observations are possible only by a long-term project
with the use of the world-largest mm-wave telescopes.

\begin{figure}
\vspace{6cm}
\caption{CO vs HI line profiles for NGC 891 (left), and
correlation between distances derived using HI and CO-line Tully 
Fisher Relations for nearby galaxies (middle) and galaxies
up to 100 Mpc (right).}
\end{figure}

On the basis of these studies, Sofue et al (1996) have conducted a long-term 
project to observe  $^{12}CO(J=1-0)$ line profiles for distant 
cluster galaxies using the Nobeyama 45-m telescope and the IRAM 30-m telescope.
Here, We summarize their result, and details will be presented in Sofue et al
(1996).
We also obtained high-quality optical imaging of detected galaxies in CO
in order to measure the inclination and magnitude using the 
Canada-France-Hawaii 3.6-m telescope.
In this paper,  we report a first-step result of the measurement
of CO line profiles in order to demonstrate that the line profiles can
be obtained in a routine work within a reasonable integration time,
and that the distances can be obtained by applying the \tfr\ relation.
We also report on optical imaging observations of several CO-detected
galaxies, and  try to obtain a possible value of the Hubble constant
for galaxies with $cz \sim 20,000$ \kms.

\subsection{CO-Line Observations}

Observations of the \co\ line of distant cluster galaxies 
were made from  1994 January 14 to 23, 1994 December 9-12,
and 1995 January 6-10 and March 13-17, using 
the 45-m telescope of the Nobeyama Radio Observatory.
We observed about fifty  galaxies, which had $cz\sim 10,000$ to 28,000 \kms
and relatively strong far-IR emission at 60 and 100 $\mu$m,
selected from the NED (NASA Extragalactic Database).
Among the galaxies, sufficient quality data have been obtained for
fifteen galaxies, and the CO line were detected in seven galaxies.
Among the seven, good CO line profiles were obtained for three
galaxies with a sufficient signal-to-noise ratio for determining the
velocity width.

\begin{figure}
\vspace{20cm}
\caption{CO line profiles for six detected galaxies and a marginal
detection (IR 14210) in channel-Ta plane.}
\end{figure}

The antenna had a HPBW of $15''$, and the aperture and  main-beam 
efficiencies were  $\eta_{\rm a}=0.35$ and $\eta_{\rm mb}=0.50$, respectively.
We used two  SIS  receivers with orthogonal polarizations
which were combined with 2048-channel acousto-optical spectrometers
of 250 MHz bandwidth with a velocity coverage of 650 (1+$z$) \kms.
After binding up every 32 channels in order to increase the signal-to-noise 
ratio, we obtained spectra with a velocity resolution of 10.2 \kms. 
The system noise temperature (SSB) was 300 to 400 K.
The calibration of the line intensity was made using an absorbing chopper 
in front of the receiver, yielding an antenna temperature (\ta), corrected 
for both the atmospheric and antenna ohmic losses.
We used an on-off switching mode, and the 
on-source total integration time was about 1-4 hours for each galaxy.
The rms noise of the resultant spectra at velocity resolution of 10 \kms 
was typically 2 mK in \ta.
The pointing of the antenna was made by observing nearby SiO maser sources 
at 43 GHz every 1 to 1.5 hr, and was typically within $\pm 3''$ during a good
weather condition which were attained for about one fourth of the allocated
observing time.

We have detected the CO-line emission for six galaxies, 
CGCG 1113.7+2936, CGCG 1448.9+1654, CGCG 1417.2+4759,
CPG 60451, NGC 6007, and IC 2846.
We obtained marginal detection for several galaxies 
including a possible detection for IRAS 14210+4829.
Observed CO spectra for the detected galaxies 
at a velocity resolution of 10 \kms\ are shown in Fig. 4.
The intensity scale used in this paper is the antenna
temperature.
We have also performed CO line observations using the IRAM 30-m
telescope, and detected CGCG 1417.2+4759.
We also show the spectrum of these galaxy taken with the 30-m telescope in 
Fig. 4.
All the obtained profiles show a typical double-peaked emission,
characteristic of a rotating ring.
From these data we measured the line width, peak antenna temperature, and line
intensity for individual galaxies.

\begin{figure}
\vspace{10cm}
\caption{
Optical images in $R$ band of galaxies CGCG 1417 (left), CGCG 1448
(middle), and CPG60451 (right). 
}
\end{figure}

\subsection{Optical Observations}

High-quality optical images of CGCG1417, CGCG1448, and CPG 60451 
at V, R and I bands were  obtained using the
Canada-France-Hawaii 3.6-m Telescope on June 30 - July 1, 1995.
Additional images have been taken with the 105-cm Schmidt telescope
of the Kiso Observatory in January 1994.
Using these data, we have determined apparent magnitude and inclination
of the galaxies.

Imaging observations were made in $V$,
$R_C$, $I_C$ bands using the Canada-France-Hawaii 3.6-m Telescope.
The CCD effective imaging area was
$2048 \times 2048$ pixels with a pixel size of 
$15\mu {\rm m} \times 15 \mu{\rm m}$.
The camera covered a sky area of $3' \times 3'$ with a resolution of 
$0.086''$ per pixel. 
The FWHM seeing size was $0.6'-- 0.8''$, which is small enough to 
determine the morphology of the observed galaxies.
The dome screen was exposed to obtain flat frames.
The large field of view gives a sufficiently large sky area around the 
galaxies to allow an accurate sky-subtraction.

Optical  $R_C$ band images are shown in Fig. 5.
The isophotal maps were fitted with ellipses to obtain the isophotal 
magnitudes and axial diameters.
To evaluate the TF relation we need the velocity width to be corrected for the 
inclination $i$ and for redshift effect into an edge-on value at the rest 
frame.
We obtain an intrinsic velocity width in the rest frame referred to the galaxy
corrected for the inclination by
$$  W_{R}^{i} = W_0~{\rm sin}~ i,$$
where $W_0$ is the rest-frame velocity width given by
$$ W_0=W_{\rm rest} = c (1+z) \Delta \nu_{\rm obs}/(115.2712~{\rm GHz}). $$
and $ \nu_{\rm obs}$ is the observed line width in frequency.
We obtain the inclination $i$ of a galaxy from the conventional formula given
by Hubble (1926) for oblate spheroid, ${\rm cos}^2i= (q^2-q_0^2) / (1-q_0^2) $
with $q=b/a$ and $q_0=c/a$, where $a$, $b$, and $c$ are the lengths of three 
axes of the spheroid.
Here, we adopt $q_0  = 0.20 $ for the present analysis.
The Galactic absorptions $A_B$,
internal absorption correction $A_i$,
and the K-correction have been applied to obtain
the corrected total magnitude 
$$m_{\rm T}^{b,i} = m_{\rm T} - A_B - A_i - K.$$

\begin{table*}[t]
\caption{Distance modulus, Distance, and Hubble ratio for the detected 
galaxies.}
\def\p{$\pm$}
\begin{tabular}{lcccccccc}
\hline\hline \\ [-5pt]
\\
Galaxy&&$ m-M$&&$D$ &&$H= <V_C/r>$ \\
      &&mag&& Mpc && \kms\ Mpc$^{-1}$ \\
\\
\hline \\ [-5pt]
 CGCG  1417   \\
 \\
 && 36.46&  \p.13&195.86 & \p11.90&110&\p7\\
\\
CGCG 1417& (IRAM 30-m)\\
 \\
 &&     37.16& \p.09&  270.55 & \p11.36&   80& \p3\\
\\
 CGCG 1448\\
 \\
 & &            36.95 &\p.08&  245.14 &\p9.28&   55&\p2\\
\\
 PGC 60451 \cr
 \\
 & &           37.33 &\p.17 &  293.21 &\p23.28 &   51 &\p4\\
\\
\hline
\end{tabular}
\end{table*}

\subsection{Distances and the Hubble Ratio}

We adopt the zero point of the TF relation given by Pierce \& Tully (1992) for
$R$ and $I$ and by Shimasaku \& Okamura (1992) for $V$ band,
and obtained the absolute magnitude.
We, then, derived distance modulae, distances, and 
Hubble ratios $V_C/r$ for individual galaxies.
Here, the recession velocity $V_C$ is the value with respect to the CMB 
rest frame.
Averaged values in the $V$, $R_C$, and $I_C$ band results are also shown
with their dispersions in Table 2.

The NRO and IRAM results are different for CGCG 1417. Since the
velocity coverage of IRAM data is wider, the IRAM result may be
more reliable, and we prefer the $H$ value of 80 for this galaxy.
For the other two galaxies, $H$ values are rather small.
However, we note that the present results are still preliminary,
and we need further observations to give conclusive values.

\section*{References}
\def\r{\hangindent=1pc  \noindent}
\small{
\r Aaronson M., Bothun G., Mould J., Shommer
   R. A., Cornell, M. E., 1986, ApJ 302, 536

\r Aaronson, M., Mould, J., \& Huchra, J. 1980, ApJ, 237, 655

\r Burton, W. B., Gordon, M. A. 1978, AA 63, 7.

\r Clemens, D. P. 1985 ApJ 295, 422.

\r Dickey J., Kazes I., 1992, ApJ 393, 530

\r Eggen, O. J., Lynden-Bell, D., and Sandage, A. 1962, ApJ 136, 748.

\r Fukugita M., Okamura S., Tarusawa K., et al., 1991, ApJ 376, 8

\r Fouqu{\'e} P., Bottinelli L., Gouguenheim L., 1990, ApJ 349, 1

\r{Genzel, R., and Townes, C.H 1987, ARAA 25, 377.}

\r Hubble, E. P. 1926, ApJ, 64, 321

\r{Ishizuki, S., Kawabe, R., Ishiguro, M., et al. 
1990 ApJ 355, 436}

\r Kraan-Korteweg R. C., Cameron L. M.,  Tammann G. A., 1988, ApJ 331, 610

\r Miyamoto, M., and Nagai, R. 1975, PASJ 27, 533.

\r{Noguchi, M. 1988, AA  203, 259. }

\r Persic, M., Salucci, P., Stel, F. 1995, MNRAS in press.

\r Pierce M. J., Tully R. B., 1988, ApJ 330, 579

\r Pierce, M. J., \&Tully, R. B. 1992, ApJ, 387, 47

\r Rubin, V. C., Ford, W. K., Thonnard, N. 1980, ApJ, 238, 471 

\r Rubin, V. C., Ford, W. K., Thonnard, N. 1982, ApJ, 261, 439 

\r Saio, H., Yoshii, Y. 1990, ApJ 363, 40.

\r  Sch\"oniger, F., Sofue, Y. 1993, AA 283, 21.

\r Shimasaku, K., \& Okamura, S. 1992, ApJ., 398, 441

\r Sofue Y., 1992, PASJ 44, L231

\r Sofue, Y. 1996a, ApJ in press.

\r Sofue, Y. 1996b, ApJ. Letter, submitted.

\r Sofue, Y., Sch\"oniger, F., Honma, M., Ichikawa, T., Wakamatsu, K., 
Kazes, I., \& Dickey, J. 1996 in preparation.

\r{S$\phi$rensen, S. -A., Matsuda, T.,  Fujimoto, M. 1976, {Ap.
 Sp. Sci. }, { 43}, 491. }

\r Staveley-Smith L., Davies R. D., 1988, MNRAS 231,  833

\r Tully B., Fisher J. R., 1977, A\&A 64, 661

\r Tully B., Fouqu{\'e} P., 1985, ApJS 58, 67

\r Wada, K., Habe, A. 1992, MNRAS 258, 82 .

}

\label{last}
\end{document}